\documentclass[runningheads,a4paper]{llncs}

\usepackage[utf8]{inputenc}
\usepackage{amssymb}
\usepackage{graphicx}
\usepackage{soul}
\usepackage{color}
\usepackage{array}
\usepackage{eurosym}
\usepackage{booktabs}
\usepackage{multirow}
\usepackage{tabularx}
\usepackage{todonotes}
\usepackage{arydshln}
\usepackage{geometry}

\usepackage{url}
\urldef{\mailsa}\path|f.dalpiaz@uu.nl,franch@essi.upc.edu,horkoff@city.ac.uk|

\begin{document}
	
	\mainmatter  
	
	\title{iStar 2.0 Language Guide}
	
	\titlerunning{iStar 2.0 Language Guide}

	\author{Fabiano Dalpiaz\inst{1}, Xavier Franch\inst{2}, and Jennifer Horkoff\inst{3}	\thanks{Endorsers: Okhaide Akhigbe, Fatma Ba\c{s}ak Aydemir, Juan Pablo Carvallo, Jaelson Castro, Luiz Marcio Cysneiros, Sepideh Ghanavati, Alicia Grubb, Giancarlo Guizzardi, Renata Guizzardi, Matthias Jarke, Alexei Lapouchnian, Tong Li, Lin Liu, Lidia López, Alejandro Maté, John Mylopoulos, Soroosh Nalchigar, Elda Paja, Angelo Susi, Juan Carlos Trujillo Mondéjar, Eric Yu, Jelena Zdravkovic.}}
	\authorrunning{F. Dalpiaz, X. Franch, and J. Horkoff}
	
	\institute{Utrecht University, The Netherlands \and
	Universitat Politècnica de Catalunya, Spain \and
	  City University London, United Kingdom \\
		\mailsa\\
		}

	\maketitle

	\setcounter{footnote}{0}

\begin{abstract}
The \textit{i*} modeling language was introduced to fill the gap in the spectrum of conceptual modeling languages, focusing on the intentional (\textit{why?}), social (\textit{who?}), and strategic (\textit{how? how else?}) dimensions.  \textit{i*} has been applied in many areas, e.g., healthcare, security analysis, eCommerce.  Although \textit{i*} has seen much academic application, the diversity of extensions and variations can make it difficult for novices to learn and use it in a consistent way. This document introduces the iStar 2.0 core language, evolving the basic concepts of \textit{i*} into a consistent and clear set of core concepts, upon which to build future work and to base goal-oriented teaching materials. This document was built from a set of discussions and input from various members of the \textit{i*} community. It is our intention to revisit, update and expand the document after collecting examples and concrete experiences with iStar 2.0. 
		
\end{abstract}

\clearpage

\section*{Version History}

\begin{table*}[!]
\centering
\begin{tabular}{|c|l|p{9cm}|}
\hline
\textit{Version} & \textit{Date} & \textit{Implemented Changes}\\
\hline
3 & June 17, 2016 & New integrity rules: at most one actor link between a pair of actors; contribution and qualification cannot connect the same two elements.\\
\hline
2 & June 3, 2016 & Fixed typos in the original version\\
\hline
1 & May 26, 2016 & -\\
\hline
\end{tabular}
\end{table*}

\clearpage
	
\section{Motivation and Overview}\label{sec:intro}
The \textit{i*} language was presented in the mid-nineties~\cite{Yu1996} as a goal- and actor-oriented modeling and reasoning framework. It consists of a modeling language along with reasoning techniques for analyzing created models. \textit{i*} was quickly adopted by the research community in fields such as requirements engineering and business modeling. Benefiting from its intentionally open nature, multiple extensions of the \textit{i*} language have been proposed (see~\cite{horkoff2014taking,horkoff2013comparison} for useful reviews), either by slightly redefining some existing constructs, by detailing some semantic issues not completely defined in the seminal proposal, or by proposing new constructs for specific domains.

This flexible use of \textit{i*} has been fruitfully employed by researchers, who were able to benefit from a consolidated modeling and reasoning approach whilst tailoring it to their needs. However, this use has also some drawbacks. The most critical is the difficulty to spread the framework outside the experts’ community:
\begin{itemize}
\item \textit{Newcomers} find it hard to learn the intricacies of the language; 
\item \textit{Educators} do not have a shared body of knowledge to teach;
\item \textit{Practitioners} are not provided with an established reference for using \textit{i*} in their projects;
\item \textit{Technology providers} cannot easily determine which are the core constructs to be implemented and the techniques to apply on top of those constructs.
\end{itemize}

As a response to the need of balancing the framework’s open nature and a possible solution to the aforementioned adoption problems, the \textit{i*} research community started an initiative to identify a widely agreed upon set of core concepts in the \textit{i*} language. The main goal is to keep open the ability to tailor the framework while agreeing on the fundamental constructs.

This document summarizes the outcomes of the first iteration of the iStar\footnote{The language is spelled iStar instead of \textit{i}* to allow better indexing through search engines.} standardization process. To clearly distinguish this core language from its predecessors, we name it \textbf{iStar 2.0}. 

The community discussed this language in several meetings and discussions starting in a dedicated one-day meeting before the ER’14 conference in Atlanta (October 2014). At the subsequent community meeting at CAiSE’15 in Stockholm (at the iStar teaching workshop, iStarT, June 2015), it was decided that a smaller group of researchers (the authors of this document) would guide the process, making concrete proposals and processing the inputs of the rest of the community. An initial draft of the core was discussed both at the iStar Workshop colocated with RE’15 (August 2015) and in another dedicated one-day meeting before ER’15 in Stockholm (October 2015). The obtained feedback has been incorporated into the document. A preliminary version was distributed among the researchers that participated in this process (December 2015), who provided a last round of comments, considered in this version (May 2016).

Each of the following five sections (Section~\ref{sec:actors}--\ref{sec:intentional-links}) addresses a particular category of language constructs as presented in most \textit{i*} sources, for example, the iStar-wiki\footnote{\url{http://istarwiki.org}}. For each category, the document lists the concepts that are included in iStar 2.0, with a definition, necessary comments, concise examples and the graphical representation. The focus of this document is on concepts and relationships; methodological possibilities are mentioned only briefly.

Several aspects are excluded from this first version of iStar but are planned for inclusion in the next versions. Among them, we mention the ontological definition of constructs (e.g., what is a goal?), visual representation (e.g., what is an effective graphical notation?), wording conventions (e.g, passive voices in goals), and methodological issues (e.g., when can a model be considered final?). 

\begin{figure}[!hc]
    \centering
    \includegraphics[width=\textwidth]{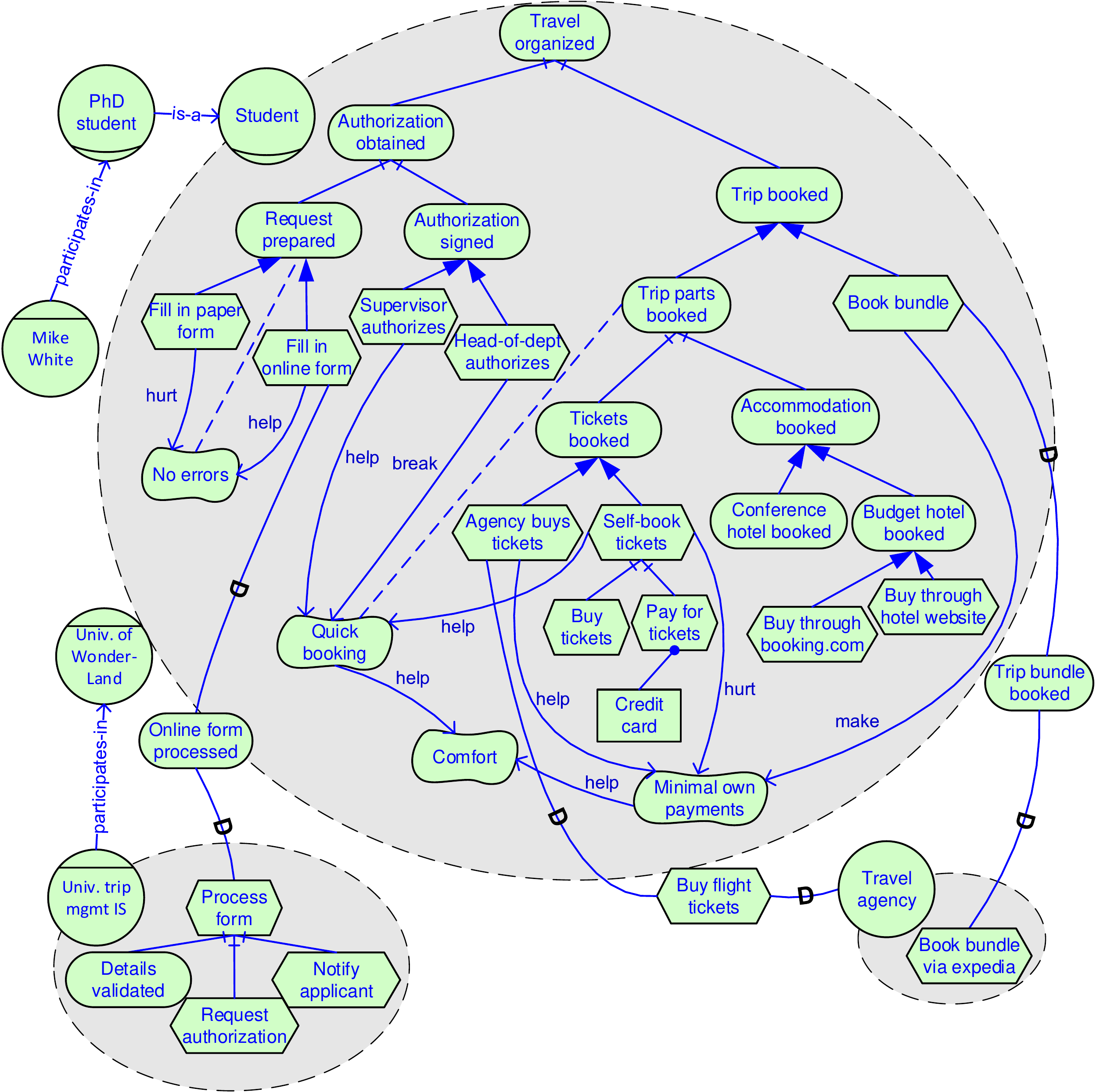}    
    \caption{A preview of the Travel Reimbursement Scenario as captured in iStar 2.0}
    \label{fig:overview}
\end{figure}

\paragraph{Illustrative example.}
We illustrate the concepts of iStar 2.0 using a running example concerning University travel reimbursement. Students must organize their travel (e.g., to conferences) and have several goals to achieve, and options to achieve them. To achieve their goals, students rely on other parties such as a Travel Agency and the university's trip management information system. We will introduce iStar concepts gradually, slowly building up the example.  In Fig.~\ref{fig:overview} we show a final view of the example in order to give readers an early idea of the capabilities of iStar 2.0.

\paragraph{Organization.} Section~\ref{sec:actors} introduces the notion of actor and distinguishes between three types of actors. Section~\ref{sec:actor-links} presents the links in iStar 2.0 for relating actors. Section~\ref{sec:intentional} describes the intentional elements that characterize the actors. Section~\ref{sec:dependencies} discusses the dependencies that socially relate the actors. Section~\ref{sec:intentional-links} explains how the intentional elements can be linked. Section~\ref{sec:views} details how to create different views of an iStar 2.0 model. Section~\ref{sec:metamodel} shows the metamodel of the language. Finally, we conclude and present future directions in Section~\ref{sec:concl}.

\section{Actors and actor types}\label{sec:actors}
Actors are central to the social modeling nature of the language. \textit{Actors} are active, autonomous entities that aim at achieving their goals by exercising their know-how, in collaboration with other actors. In the iStar 2.0 language, two types of actors are distinguished:
\begin{itemize}
\item \textit{Role}: an abstract characterization of the behavior of a social actor within some specialized context or domain of endeavor. Examples are: Student, PhD Student.
\item \textit{Agent}: an actor with concrete, physical manifestations, such as a human individual, an organization, or a department. Examples are: Travel agency, PhD Student, University of Wonderland, Mike White.
\end{itemize}

Whenever distinguishing the type of actor is not relevant, either because of the scenario-at-hand or the modeling stage, the notion of generic \textit{actor}---without specialization---can be used in the model. For example, we can denote Travel agency as an actor to say that we do not know yet whether it is a specific agency (agent) or a characterization of the travel agency role. 

\begin{figure}[!hc]
    \centering
    \includegraphics[width=0.4\textwidth]{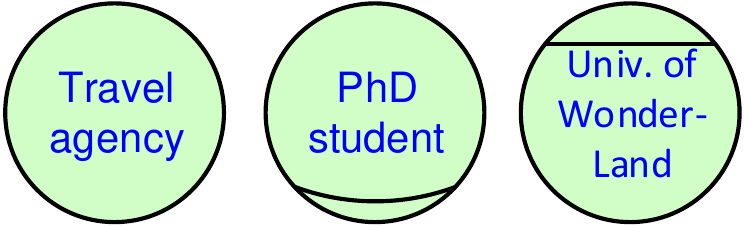}    
    \caption{Examples of actor, role and agent}
    \label{fig:actors}
\end{figure}

Actors are represented graphically as circles. In the case of agent, a straight line is added in the top part of the actor circle. For a role, a curved line is added in the lower part. The graphical notation follows the mnemonic guidelines that the original \textit{i*} adopts\footnote{These symbols are stylized depictions of a person wearing a hat and viewed from different angles. The Agent symbol is the frontal view where the name of the agent appears on the face. The Role symbol is an overhead view so that the label on the hat is visible.}. Fig.~\ref{fig:actors} illustrates the notation.

\begin{figure}[!hc]
    \centering
    \includegraphics[width=0.22\textwidth]{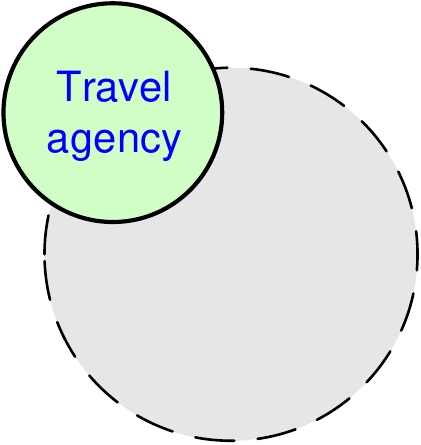}    
    \caption{Example of actor boundary}
    \label{fig:boundary}
\end{figure}

Actors' intentionality is made explicit through the \textit{actor boundary}, which is a graphical container for their intentional elements (see Section~\ref{sec:intentional}) together with their interrelationships (see Section~\ref{sec:intentional-links}). Fig.~\ref{fig:boundary} shows the graphical representation of an actor boundary; elements and relationships will appear inside the grey area.

\section{Actors association links}\label{sec:actor-links}
Actors are often interrelated. In iStar 2.0, this is captured via actor links that define/describe these relationships. Actor links are binary, linking a single actor to a single other actor. Two different types of actor links have been defined:
\begin{itemize}
\item \textit{is-a}: represents the concept of generalization / specialization in iStar 2.0. Only roles can be specialized into roles, or general actors into general actors. For instance, a PhD student (role) can be defined as a specialization of a Student (another role). Agents cannot be specialized via is-a, as they are concrete instantiations (e.g., Mike White cannot be another agent).
\item \textit{participates-in}: represents any kind of association, other than generalization / specialization, between two actors. No restriction exists on the type of actors linked by this association. Depending on the connected elements, this link takes different meanings. Two typical situations are the following:
	\begin{itemize}
	\item When the source is an agent and the target is a role, this represents the \textit{plays} relationship, i.e., an agent plays a given role. For instance, Mike White plays the role of PhD student.
	\item When the source and the target are of the same type, this will often represent the \textit{part-of} relationship. For instance, the University trip management information system is part of the University of Wonderland.
	\end{itemize}
	Every actor can \textit{participate-in} multiple other actors.
\end{itemize}
\begin{figure}[!hc]
    \centering
    \includegraphics[width=0.4\textwidth]{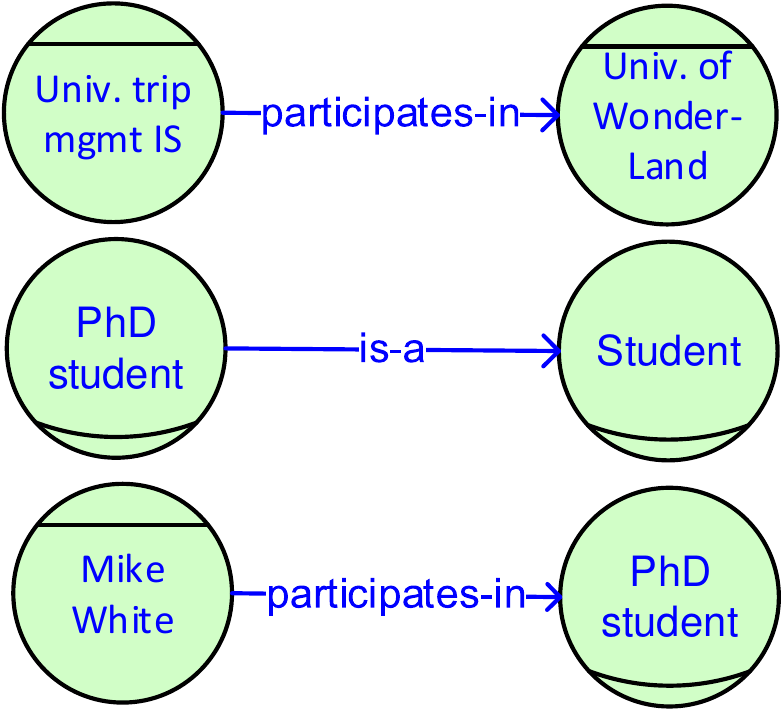}    
    \caption{Examples of actor association links}\vspace*{-0.5cm}
    \label{fig:actor-association}
\end{figure}
Actor association links are represented using arrows in the diagram. The arrowhead identifies the target (the participated actor or the superclass, respectively). A label identifying the type of link must be included. Fig.~\ref{fig:actor-association} shows the graphical representation of the three examples mentioned above.

\section{Intentional elements}\label{sec:intentional}
\begin{figure}[!hc]
    \centering
    \includegraphics[width=0.55\textwidth]{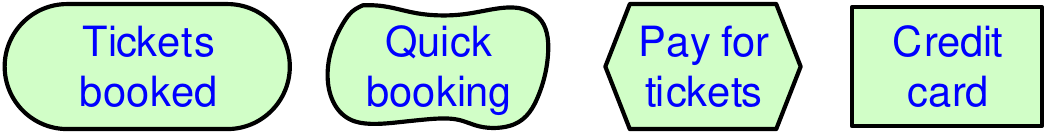}    
    \caption{Examples of intentional elements}
    \label{fig:intel}\vspace*{-1cm}
\end{figure}

Intentional elements are the things actors want. As such, they model different kinds of requirements and are central to the iStar 2.0 language. An intentional element appearing inside the boundary of an actor denotes something that is desired or wanted by that actor. An intentional element can also appear outside of actor boundaries, as part of a dependency relationship between two actors (see Section~\ref{sec:dependencies}). In this section, we focus on the former case: elements inside actor boundaries. The following elements are included in the language:
\begin{itemize}
\item \textit{Goal}: a state of affairs that the actor wants to achieve and that has clear-cut criteria of achievement.
\item \textit{Quality}: an attribute for which an actor desires some level of achievement. For example, the entity could be the system under development and a quality its performance; another entity could be the business being analyzed and a quality the yearly profit. The level of achievement may be defined precisely or kept vague. Qualities can guide the search for ways of achieving goals, and also serve as criteria for evaluating alternative ways of achieving goals\footnote{iStar 2.0 departs from the original goal / softgoal dichotomy, which distinguishes these concepts based on the existence of a clear-cut metric for satisfaction, and from the Non-Functional/Functional Requirement distinction, as the use of this distinction varied in practice. By including qualities, which can be either ``soft'' or ``hard'' using \textit{i*} terminology, and by including the qualifies relationship between qualities and goals (Section~\ref{sec:qualif}), we clarify the relationships between goals and qualities.}.  
\item \textit{Task}: represents actions that an actor wants to be executed, usually with the purpose of achieving some goal.
\item \textit{Resource}: A physical or informational entity that the actor requires in order to perform a task.
\end{itemize}
Goals are graphically represented as ovals, while qualities are represented as more curved cloud-like shapes. Tasks are represented as hexagons to highlight their more structured definition in terms of a process to be followed. Resources are represented as rectangles. Fig.~\ref{fig:intel} shows examples of the intentional elements.

\section{Social dependencies}\label{sec:dependencies}
\textit{Dependencies} represent social relationships in iStar 2.0. This, along with the assumption that actors can be human, organizations, technical systems (hardware, software), or any combination thereof, makes iStar 2.0 a socio-technical modeling language. A dependency is defined as a relationship with five arguments:
\begin{itemize}
\item \textit{depender} is the actor that depends for something (the dependum) to be provided;
\item \textit{dependerElmt} is the intentional element within the depender’s actor boundary where the dependency starts from, which explains why the dependency exists;
\item \textit{dependum} is an intentional element that is the object of the dependency;
\item \textit{dependee} is the actor that should provide the dependum;
\item \textit{dependeeElmt} is the intentional element that explains how the dependee intends to provide the dependum.
\end{itemize}
\begin{figure}[!hc]
    \centering
    \includegraphics[width=0.8\textwidth]{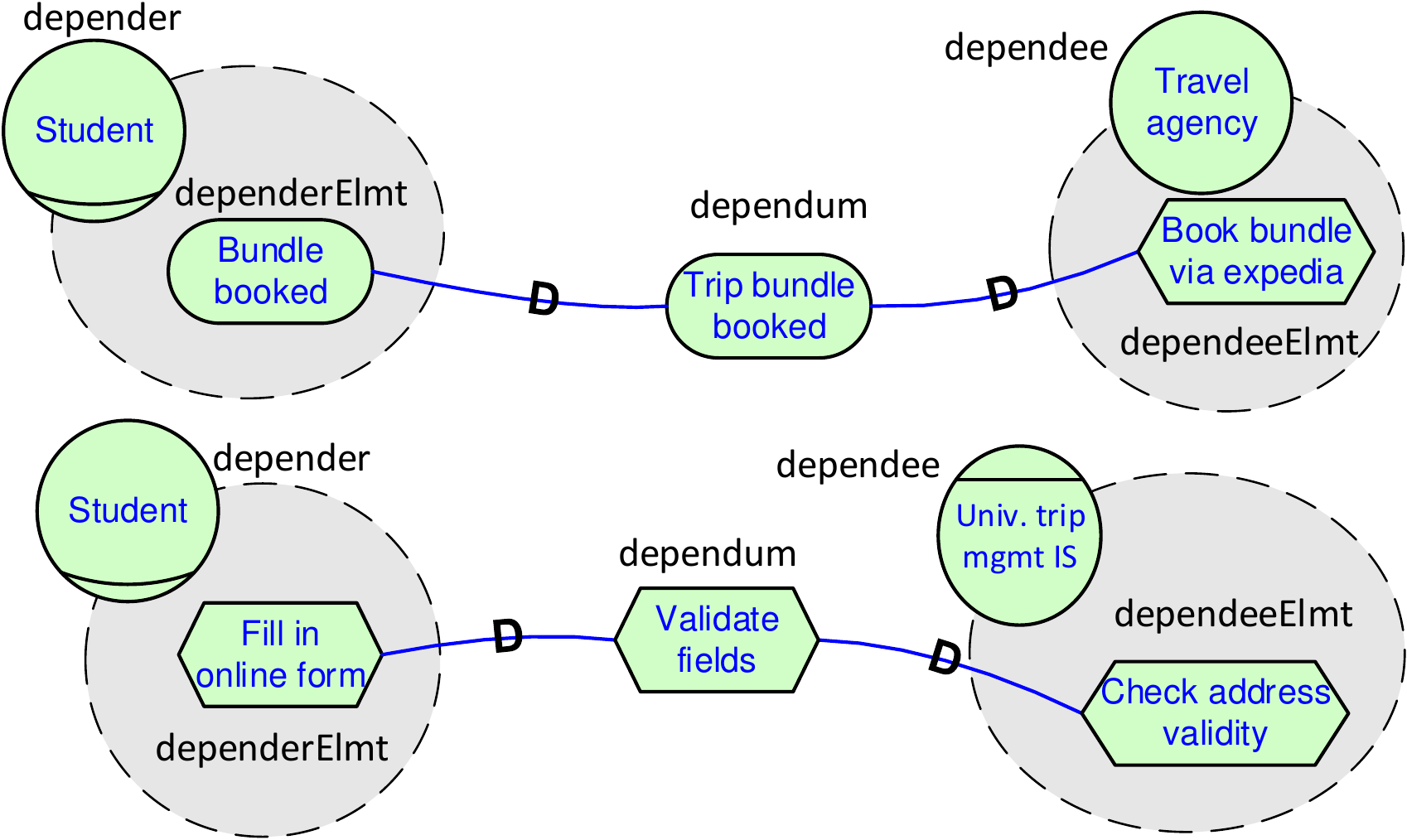}    
    \caption{Examples of dependencies}
    \label{fig:dep}
\end{figure}
Dependencies (illustrated in Fig.~\ref{fig:dep}) link the dependerElmt within the depender actor to the dependum, outside actor boundaries, to the dependeeElmt within the dependee actor. The link is drawn with a ``D'' symbol indicating direction, with the D acting as an arrowhead ``$>$'', pointing from dependerElmt to dependum to dependeeElmt.

\begin{figure}[!hc]
    \centering
    \includegraphics[width=0.8\textwidth]{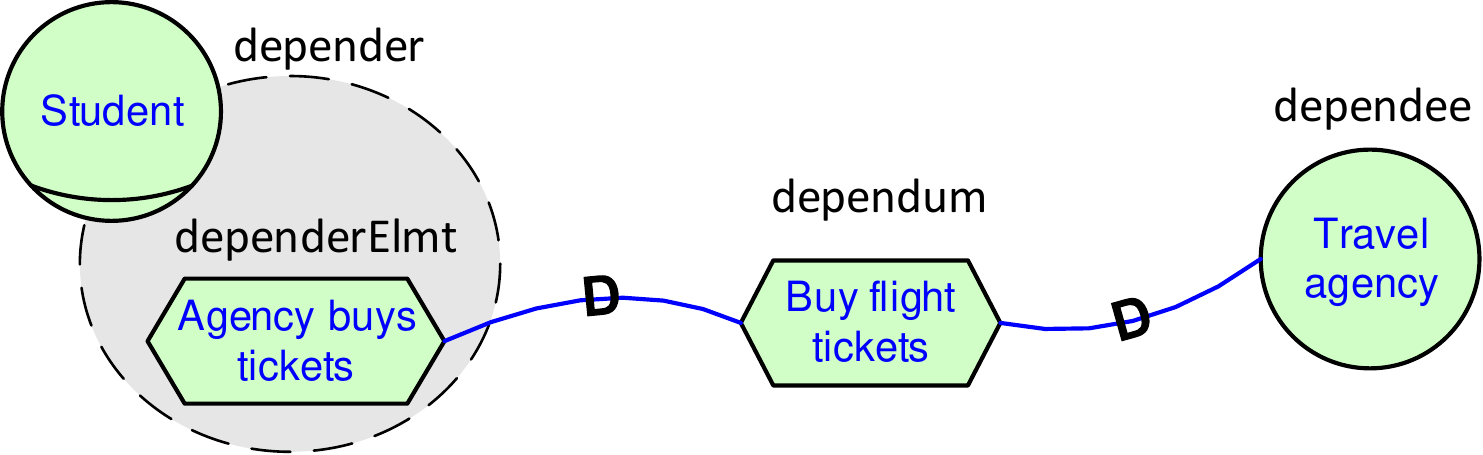}    
    \caption{Example dependency with dependeeElmt omitted}
    \label{fig:dep2}
\end{figure}

Both the dependerElmt and the dependeeElmt can be omitted. This optionality is used when creating an initial Strategic Dependency view (see Section~\ref{sec:views}), or to support expressing partial knowledge, e.g., when the ``why'' (dependerElmt) or the ``how''; (dependeeElmt) of the dependency are unknown. If both are omitted, the dependency links the depender actor to the dependee actor through the dependum. It is also possible to specify only one of them. See Fig.~\ref{fig:dep2} for an example where the dependeeElmt is omitted.

The type of the dependum specializes the semantics of the relationship:
\begin{itemize}
\item Goal: the dependee is expected to achieve the goal, and is free to choose how;
\item Quality: the dependee is expected to sufficiently satisfy the quality, and is free to choose how;
\item Task: the dependee is expected to execute the task in a prescribed way;
\item Resource: the dependee is expected to make the resource available to the depender.
\end{itemize}
This way, different dependency types indicate different degrees of freedom afforded by the depender to the dependee, with qualities and goals allowing the highest degree of freedom, tasks medium, and resource the lowest.  

\subsection*{Rules and restrictions}
When a depender depends on the dependee for its dependerElmt, the depender cannot or chooses not to satisfy/perform/have the dependerElmt on its own. Thus, the dependerElmt cannot be refined or contributed to.

It is possible that one or more of the dependerElmt, dependum, and dependeeElmt have the same element name. In this case, each of these elements is nevertheless distinct, as they reflect the separate viewpoints of the depender, the relationship between the two actors, and of the dependee respectively. 

Dependency relationships should not share the same dependum, as each dependum is a conceptually different element; in some cases, a dependum in one dependency is achieved, but is not achieved in another dependency, even if the dependums may have the same name. In other words, an actor cannot depend on more than one actor for the same dependum, or two actors cannot depend on the same dependum from an actor.

\section{Intentional element links}\label{sec:intentional-links}
There are four types of links between intentional elements: \textit{refinement}, \textit{needed-by}, \textit{contribution} and \textit{qualification}. These are described in the following sub-sections and are summarized in Table~\ref{tab:intlinks}.

\vspace*{-0.2cm}
\begin{table}[htc]
\centering
\caption{Links between intentional elements: overview}\label{tab:intlinks}
\begin{tabular}{|c|c|c;{1pt/3pt}c;{1pt/3pt}c;{1pt/3pt}c|}
\hline
\multicolumn{2}{|c|}{} & \multicolumn{4}{c|}{\textbf{Arrowhead pointing to}}\\\cline{3-6}
\multicolumn{2}{|c|}{} & \multicolumn{1}{c|}{\textit{Goal}} &  \multicolumn{1}{c|}{\textit{Quality}} &  \multicolumn{1}{c|}{\textit{Task}} & \textit{Resource}\\ 
\hline
\multirow{4}{*}{\textbf{Link starts from}} & \textit{Goal} & Refinement & Contribution & Refinement & n/a \\ \cline{2-2}\cdashline{3-6}[1pt/3pt]
& \textit{Quality} & Qualification & Contribution & Qualification & Qualification\\ \cline{2-2}\cdashline{3-6}[1pt/3pt]
& \textit{Task} & Refinement & Contribution & Refinement & n/a \\\cline{2-2}\cdashline{3-6}[1pt/3pt]
& \textit{Resource} & n/a & Contribution & NeededBy & n/a \\
\hline
\end{tabular}
\end{table}

\subsection{Refinement}\label{sec:refin}
To promote ease of adoption, iStar 2.0 features a generic relationship called \textit{refinement} that links goals and tasks hierarchically. Refinement is an n-ary relationship relating one parent to one or more children. An intentional element can be the parent in at most one refinement relationship.

Two types of refinement exist---and apply to any kind of parent (goal or task)---that define the logical operator that relates the parent with the children:
\begin{itemize}
\item \textit{AND}: the fulfillment of all the $n$ children ($n\geq 2$) makes the parent fulfilled;
\item \textit{Inclusive OR}: the fulfillment of at least one child makes the parent fulfilled. This relationship allows for a single child.
\end{itemize}

A parent can only be AND-refined or OR-refined, not both simultaneously.  Depending on the connected elements, refinement takes different meanings:
\begin{itemize}
\item If the parent is a \textit{goal}:
	\begin{itemize}
	\item In the case of AND, a \textit{child goal} is a sub-state of affairs that is part of the parent goal, while a \textit{child task} is a sub-task that must be fulfilled;
	\item In the case of OR, a \textit{child task} is a particular way (a ``means") for fulfilling the parent goal (the ``end"), while a \textit{child goal} is a sub-goal that can be achieved for fulfilling the parent goal;
	\end{itemize}
\item If the parent is a \textit{task}:
	\begin{itemize}
	\item In the case of AND, a \textit{child task} is a sub-task  that is identified as part of the parent task, while a \textit{child goal} is a goal that is uncovered by analyzing the parent task;
	\item In the case of OR, a \textit{child goal} is a goal whose existence that is uncovered by analyzing the parent task which may substitute for the original task, while a \textit{child task} is a way to execute the parent task.
	\end{itemize}
\end{itemize}
Refinement relationships do not imply a strictly top-down process, they can be built from the bottom-up, top-down, or via a mixed approach.  

\begin{figure}[!hc]
    \centering
    \includegraphics[width=\textwidth]{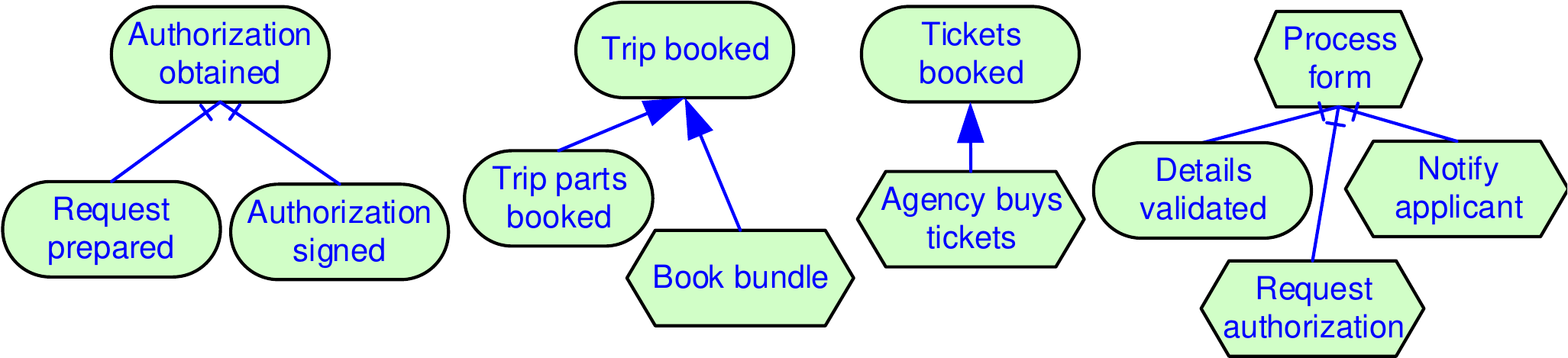}    
    \caption{Examples of refinement links}
    \label{fig:refinement}
\end{figure}

Graphically\footnote{As mentioned in Section~\ref{sec:intro}, an accurate study of the graphical definition of refinement (and other relationships) is left to future versions of the language.}, \textit{refinement} is expressed as a set of links directed from the sub-elements to the parent element. We employ a T-shaped arrowhead to denote AND-refinement, and a solid arrow directed towards the parent to represent OR-refinement (we use the original \textit{i*} symbol). See Fig.~\ref{fig:refinement} for examples.

The first refinement shows a goal decomposed into two sub-goals that are both necessary (\textit{AND-refinement}) to achieve the parent. The second refinement shows alternatives (\textit{OR-refinement}): to book a trip, either the parts are booked, a bundle is booked, or both alternatives are chosen; while the former sub-element is a sub-state of affairs to achieve (a \textit{goal}), the latter sub-element is a concrete set of actions to execute (a \textit{task}). The third refinement exemplifies the existence of a single alternative. The fourth refinement shows the uncovering of goals while analyzing tasks: the goal “Details validated” is in the model because of the task “Process form”.

\subsection{NeededBy}\label{sec:neededby}
\begin{figure}[!]
    \centering
    \includegraphics[width=0.12\textwidth]{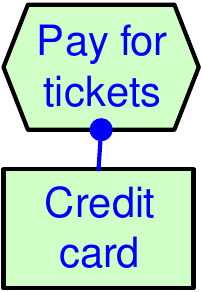}    
    \caption{Example of NeededBy relationship}\vspace*{-0.8cm}
    \label{fig:neededby}
\end{figure}

The \textit{NeededBy} relationship links a task with a resource and it indicates that the actor needs the resource in order to execute the task. This relationship does not specify what is the reason for this need: consumption, reading, modification, creation, etc. Graphically, NeededBy is represented as an arrow with a circle arrowhead directed towards the task, as shown in Fig.~\ref{fig:neededby}.

\subsection{Contribution}\label{sec:contrib}
\textit{Contribution links} represent the effects of intentional elements on qualities, and are essential to assist analysts in the decision-making process among alternative goals or tasks. Contribution links lead to the accumulation of evidence for qualities. We talk of qualities being \textit{fulfilled} or \textit{satisfied}, having sufficient positive evidence, or being \textit{denied}, having strong negative evidence.

Contributions are defined as relationships from a source intentional element to a target quality, and having one of the following types:
\begin{itemize}
\item \textit{Make}: The source provides sufficient positive evidence for the satisfaction of the target.
\item \textit{Help}: The source provides weak positive evidence for the satisfaction of the target.
\item \textit{Hurt}: The source provides weak evidence against the satisfaction (or for the denial) of the target. 
\item \textit{Break}: The source provides sufficient evidence against the satisfaction (or for the denial) of the target.
\end{itemize}

\begin{figure}[!hc]
    \centering
    \includegraphics[width=.75\textwidth]{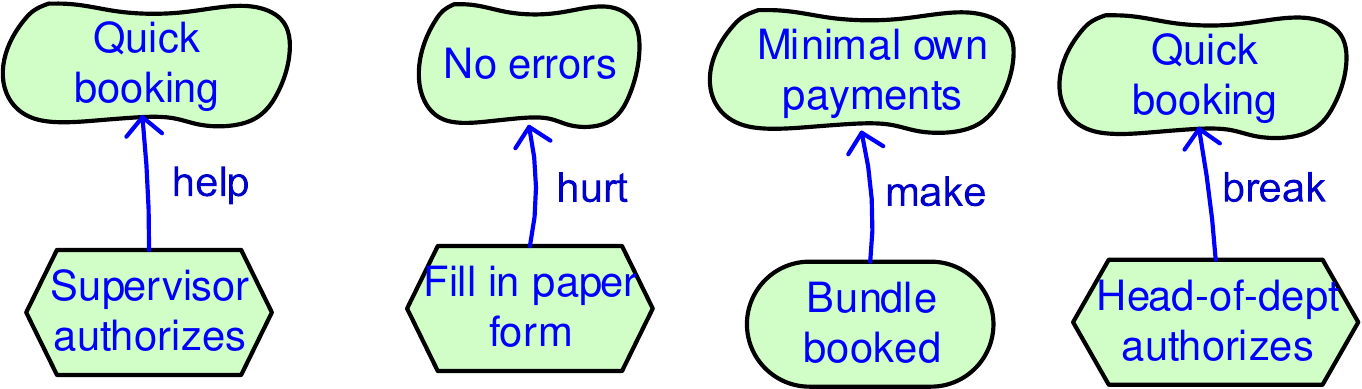}    
    \caption{Example of contribution links}
    \label{fig:contrib}
\end{figure}

Contributions are represented graphically as solid arrows with a text label that indicates the contribution type (see Fig.~\ref{fig:contrib}). While the examples show contributions starting from goals and tasks, it is also possible to initiate contributions from resources and qualities.

\subsection{Qualification}\label{sec:qualif}
The \textit{qualification} relationship relates a quality to its subject: a task, goal, or resource. For example, the quality “Quick booking” refers to the goal “Trip parts booked”, it qualifies how the operation or function of this goal should be achieved. Similarly, the quality “No errors” refers to errors possibly created while fulfilling the goal “Request prepared”, elaborating on how this goal might be achieved. Qualities are not necessarily attached to other elements through a qualification link. For example, “Avoid own payments” can be a standalone quality that does not qualify any other element, particularly if a task or goal concerning making one's own payments is not present in the model.

\begin{figure}[!hc]
    \centering
    \includegraphics[width=.22\textwidth]{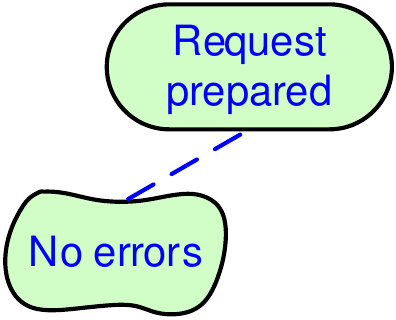}    
    \caption{Example of qualification link}
    \label{fig:qualif}
\end{figure}

Placing a qualification relationship expresses a desired quality over the execution of a task, the achievement of the goal, or the provision of the resource. For example, in Fig.~\ref{fig:qualif}, saying that “No errors” qualifies “Request prepared” means that the preparation of a request should be achieved in such a way that it leads to no errors. The qualification relationship is represented graphically via a dotted line connecting  the element that is qualifies.

\section{Model views}\label{sec:views}
When using iStar 2.0, the analyst creates a \textit{model}. Such model can be visualized via multiple perspectives or \textit{model views} (see also~\cite{Dalpiaz2016}). We specifically introduce three views that stem from the original \textit{i*} proposal and some extensions: the Strategic Rationale view, the Strategic Dependency view, and the Hybrid view. 

\paragraph{Strategic Rationale (SR).} The SR view shows all of the detail captured in the model, including actors, dependencies, actor association links, and the internal details of each actor. Modelers can view the strategic rationale inside each of the actors in the model. An SR view for the travel reimbursement scenario was shown at the beginning of this document, in Fig.~\ref{fig:overview}.

\begin{figure}[!h]
    \centering
    \includegraphics[width=.8\textwidth]{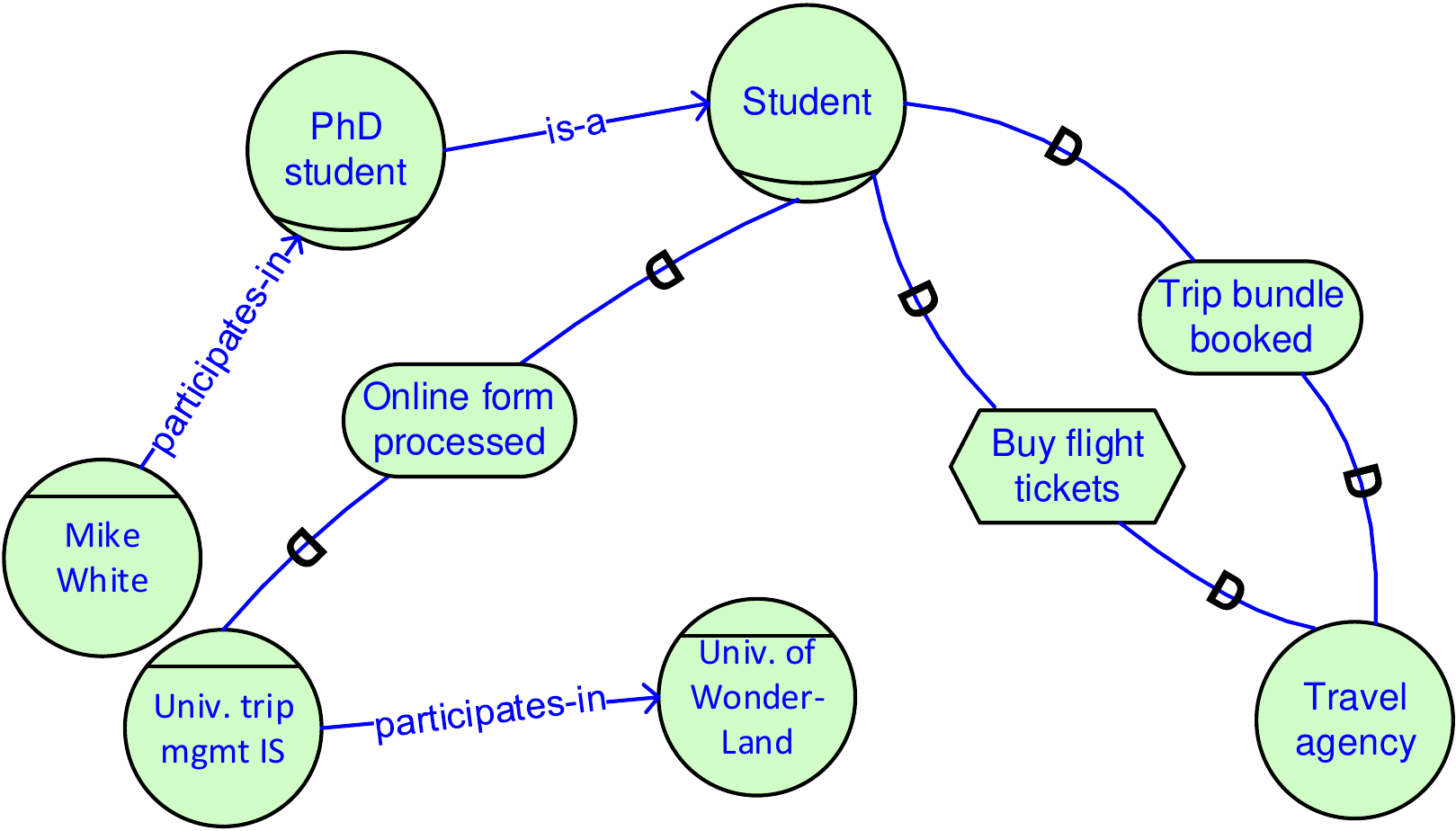}    
    \caption{An SD view of the Travel Reimbursement scenario}
    \label{fig:sdview}
\end{figure}

\paragraph{Strategic Dependency (SD).} The SD view shows each actor in the model, the actor association links, and the dependency relationships from depender to dependum to dependee. Fig.~\ref{fig:sdview} shows an SD view of the travel reimbursement scenario.

\begin{figure}[!hc]
    \centering
    \includegraphics[width=\textwidth]{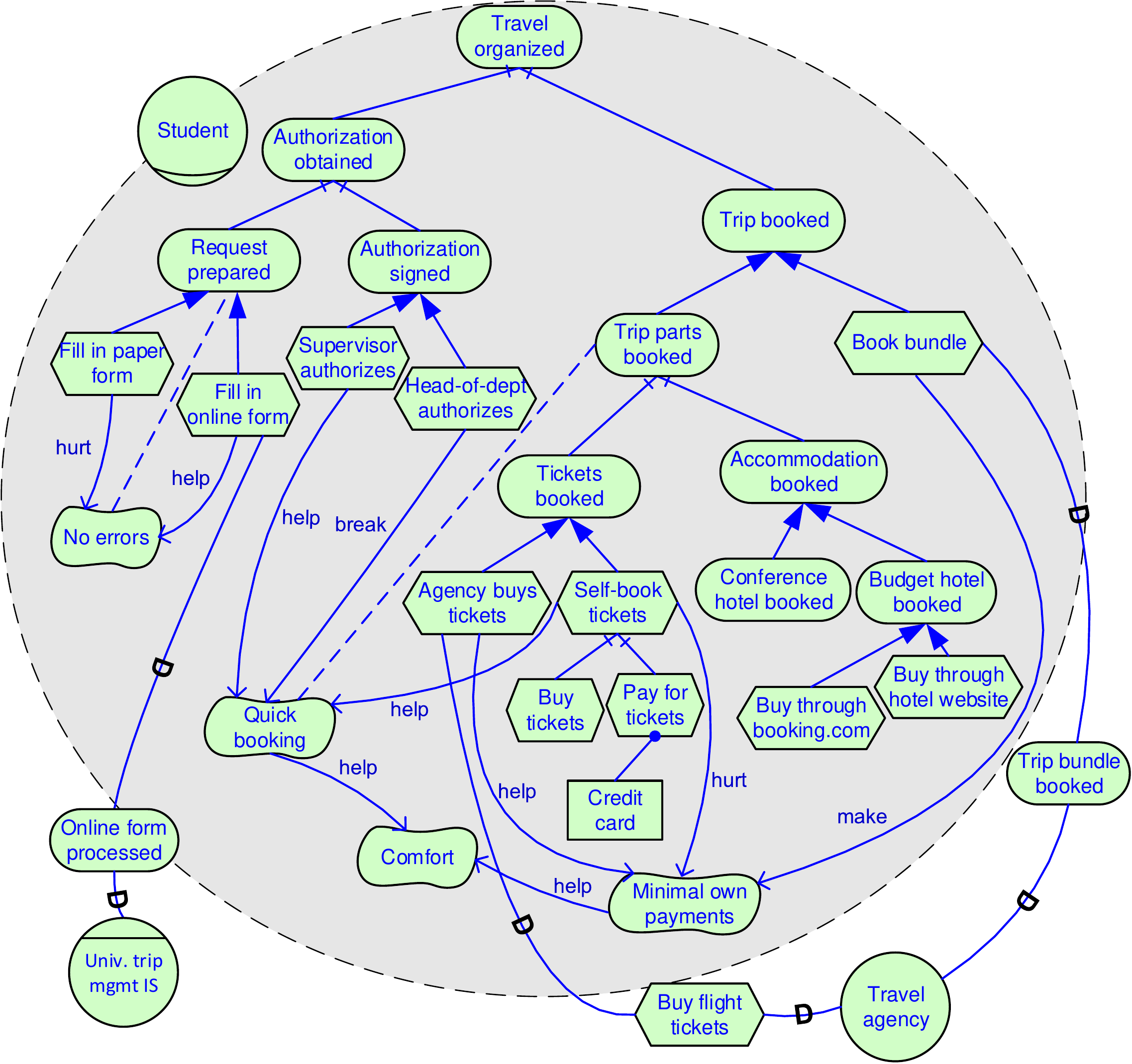}    
    \caption{A hybrid SD/SR view of the Travel Reimbursement scenario}
    \label{fig:hybrid}
\end{figure}

\paragraph{Hybrid SD/SR.} It is often useful to combine SD/SR views where some of the actors are open, but not all, focusing on the strategic rationale of a particular set of actors, and the actor links are hidden. This view is illustrated in Fig.~\ref{fig:hybrid}.

Further useful views can be defined as needed, for example, the \textit{actor view}, showing only actors and actor links, or a \textit{functional view}, hiding all qualities, contribution and restriction links.

\section{Metamodel}\label{sec:metamodel}
\begin{figure}[!hc]
    \centering
    \includegraphics[width=\textwidth]{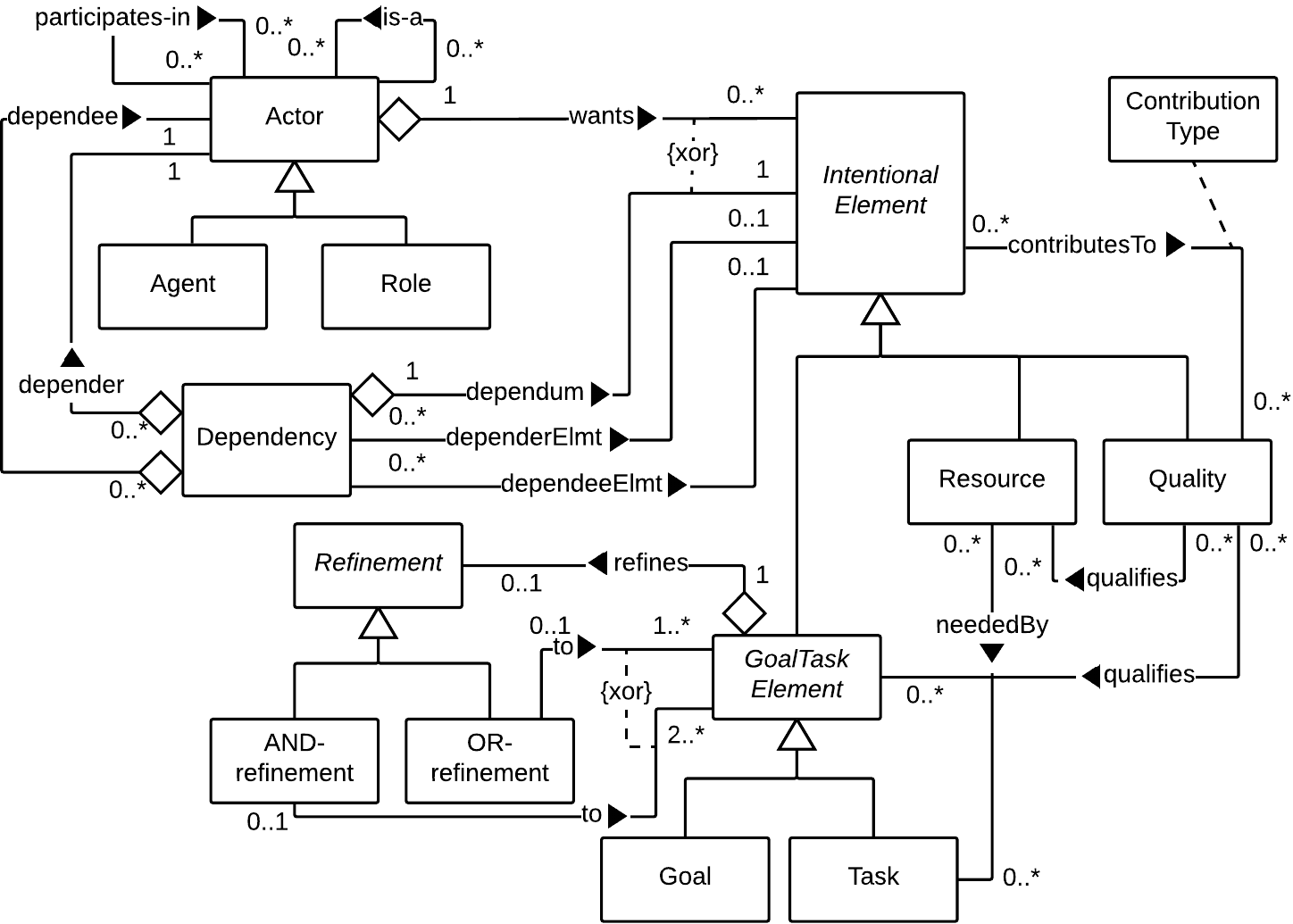}    
    \caption{Metamodel of iStar 2.0}
    \label{fig:metamodel}
\end{figure}

The metamodel for iStar 2.0 is shown in Fig.~\ref{fig:metamodel}.  The concepts and relationships in the metamodel have been explained and illustrated in the previous sections. 
We describe a number of integrity constraints that explain more detailed constraints on the models that the iStar 2.0 metamodel allows:
\begin{itemize}
\item The \textit{is-a} relationship applies only between pairs of roles or pairs of actors;
\item There should be no \textit{is-a} cycles;
\item There should be no \textit{participates-in} cycles;
\item A pair of actors can be linked by at most one actor link: it is not possible to connect two actors via both \textit{is-a} and \textit{participates-in};
\item In a dependency $D$, if the \textit{dependerElmt} $x$ exists, then the actor that wants $x$ is the same actor that is $D$'s \textit{depender};
\item In a dependency $D$, if the \textit{dependeeElmt} $y$ exists, then the actor that wants $y$ is the same actor that is $D$'s \textit{dependee};
\item The \textit{depender} and \textit{dependee} of a dependency should be different actors;
\item For a dependency, if a \textit{dependerElmt} $x$ exists, then $x$ cannot be refined or contributed to;
\item The refinement relationship should not lead to refinement cycles (e.g., $G$ OR-refined to $G_1$ and $G_1$ OR-refined to $G$, $G$ OR-refined to $G$, etc.);
\item The relationships between intentional elements (\textit{contributesTo}, \textit{qualifies}, \textit{neededBy}, \textit{refines}) apply only to elements that are wanted by the same actor;
\item An intentional element and a quality can be linked by either a \textit{contributesTo} relationship or a \textit{qualifies} relationship, but not by both;
\item It is not possible for a quality to contribute to itself. 
\end{itemize}
Finally, note that some classes in the metamodel are abstract (\textit{GoalTask Element}, \textit{Intentional Element}, \textit{Refinement}) as they are used to group together (via specialization) concrete classes that share some characteristics. On the other hand, \textit{Actor} is also specialized but it is a concrete class, thereby denoting that actors can be instantiated without further specialization.

\section{Conclusion and Outlook}\label{sec:concl}
This document presented the results of the iStar standardization process that has led to the definition of the iStar 2.0 language. Supported by the research community of iStar, we promote the adoption of iStar 2.0 for educational and training purposes. To such extent, one of the next steps will be the creation of teaching materials that can be readily used to teach iStar 2.0.

This document does not conclude the standardization process. Following iterative design principles, we encourage the entire community to assist us in the conduction of studies about the ease of use, the adequacy for teaching, the expressiveness, the graphical notation, and the automated reasoning techniques that can support iStar 2.0.

Although our efforts in reconciling the numerous viewpoints of the community, we are well aware that there is still room for improvement. Therefore, we warmly welcome your feedback concerning the chosen primitives, the graphical notation, typographic errors, etc. We especially welcome examples of models that are created using iStar 2.0, which are especially helpful to pinpoint the aspects that can be improved.

Contact us by sending an e-mail or post public comments on the iStar 2.0 website: \url{https://sites.google.com/site/istarlanguage/}.


\bibliographystyle{plain}
\bibliography{biblio}

\begin{thebibliography}{1}

\bibitem{Dalpiaz2016}
Fabiano Dalpiaz, Elda Paja, and Paolo Giorgini.
\newblock {\em {Security Requirements Engineering: Designing Secure
  Socio-Technical Systems}}.
\newblock MIT Press, 1 edition, 2016.

\bibitem{horkoff2014taking}
Jennifer Horkoff, Tong Li, Feng-Lin Li, Mattia Salnitri, Elsa Cardoso, Paolo
  Giorgini, John Mylopoulos, and Jo{\~a}o Pimentel.
\newblock Taking goal models downstream: a systematic roadmap.
\newblock In {\em International Conference on Research Challenges in
  Information Science}, pages 1--12. IEEE, 2014.

\bibitem{horkoff2013comparison}
Jennifer Horkoff and Eric Yu.
\newblock Comparison and evaluation of goal-oriented satisfaction analysis
  techniques.
\newblock {\em Requirements Engineering}, 18(3):199--222, 2013.

\bibitem{Yu1996}
Eric Siu-Kwong Yu.
\newblock {\em {Modelling strategic relationships for process reengineering}}.
\newblock PhD thesis, University of Toronto, 1996.

\end{thebibliography}
	
\end{document}